\documentclass[aps,prb,twocolumn,superscriptaddress]{revtex4}
\usepackage{graphicx,epstopdf}
\usepackage{amsmath,amssymb}
\newcommand{\pdag}{{\phantom{\dagger}}}

\newcommand{\pcco}{Pr$_{2-x}$Ce$_x$CuO$_{4-\delta}$}
\newcommand{\ncco}{Nd$_{2-x}$Ce$_x$CuO$_{4-\delta}$}

\begin{document}

% Use the \preprint command to place your local institutional report
% number in the upper righthand corner of the title page in preprint mode.
% Multiple \preprint commands are allowed.
% Use the 'preprintnumbers' class option to override journal defaults
% to display numbers if necessary
%\preprint{}

%Title of paper
\title{Nernst effect in the electron-doped cuprates}

% repeat the \author .. \affiliation  etc. as needed
% \email, \thanks, \homepage, \altaffiliation all apply to the current
% author. Explanatory text should go in the []'s, actual e-mail
% address or url should go in the {}'s for \email and \homepage.
% Please use the appropriate macro foreach each type of information

% \affiliation command applies to all authors since the last
% \affiliation command. The \affiliation command should follow the
% other information
% \affiliation can be followed by \email, \homepage, \thanks as well.
\author{Andreas Hackl}
\affiliation{Department of Physics, Harvard University, Cambridge, Massachusetts 02138, USA}

\affiliation{Institut f\"ur Theoretische Physik, Universit\"at zu K\"oln,
Z\"ulpicher Stra\ss e 77, 50937 K\"oln, Germany}

\author{Subir Sachdev}
%\email[]{Your e-mail address}
%\homepage[]{Your web page}
%\thanks{}
%\altaffiliation{}
\affiliation{Department of Physics, Harvard University, Cambridge, Massachusetts 02138, USA}

%Collaboration name if desired (requires use of superscriptaddress
%option in \documentclass). \noaffiliation is required (may also be
%used with the \author command).
%\collaboration can be followed by \email, \homepage, \thanks as well.
%\collaboration{}
%\noaffiliation

\date{\today}

\begin{abstract}
We calculate the normal state Nernst signal in the cuprates resulting
from a reconstruction of the Fermi surface due to spin density wave order.
An order parameter consistent with the reconstruction of the Fermi surface 
detected in electron-doped materials is shown to sharply enhance the Nernst
signal close to optimal doping. Within a semiclassical treatment,
the obtained magnitude and position of the enhanced Nernst signal agrees
with Nernst measurements in electron-doped cuprates. Our result
is mainly caused by the role of Fermi surface geometry under  
influence of a spin density wave gap. We discuss also possible roles of 
short-ranged magnetic order in the normal state Nernst effect
and the Fermi surface reconstruction observed by photoemission spectroscopy.
\end{abstract}

% insert suggested PACS numbers in braces on next line
% insert suggested keywords - APS authors don't need to do this
%\keywords{}

%\maketitle must follow title, authors, abstract, \pacs, and \keywords
\maketitle

% body of paper here - Use proper section commands
% References should be done using the \cite, \ref, and \label commands

\section{\label{intro} Introduction}

The Nernst effect has emerged as one of the key probes of the enigmatic
underdoped phase of the cuprate high temperature superconductors.
In the hole-doped case, observations \cite{Xu2000} of a strongly enhanced Nernst signal
at temperatures ($T$) well above the superconducting $T_c$
have been interpreted \cite{Xu2000,vortex} using a picture of a liquid of vortices in the superconducting order.
However, there have also been suggestions \cite{bbbss} that spin/charge density wave correlations of the
vortex liquid are important. In particular, a model of fluctuations associated with the quantum phase transition (QPT)
to the ordered stripe state at hole doping $\delta=1/8$ has been argued \cite{markus} to have a Nernst
response qualitatively similar to the observations.

In this paper, we focus on the electron-doped cuprates, where the situation appears 
simpler. The only observed order (apart from superconductivity)
is a spin density wave (SDW) which remains commensurate at the $(\pi, \pi)$ wavevector
(in the Brillouin zone of a square lattice of unit lattice spacing). 
The Nernst effect, being unmeasurable small in nearly all metals, has also been found to 
be anomalously large near optimal doping in the normal state of electron-doped 
cuprates \cite{Fournier1997,Greene2007}. We will show here that this large Nernst signal
can be understood in a theory of Fermi surface reconstruction associated with
the QPT involving onset of SDW order.

The large normal state Nernst signals found in \pcco\ (PCCO) \cite{Greene2007} 
upon Ce doping, and in \ncco\ (NCCO) upon oxygen doping, \cite{Fournier1997}
have been attributed to the existence of two types of carriers, which 
avoid the Sondheimer cancellation of the Nernst signal expected in single 
carrier systems. Indeed, angle resolved photoemission  spectroscopy (ARPES) experiments 
on NCCO found both electron- and hole-like Fermi pockets near optimal doping 
\cite{Armitage2002}. In the underdoped region, only small electron-like pockets 
remain, while in the overdoped region, only a large hole-like pocket centered at 
$(\pi,\pi)$ was found \cite{Matsui2007}. These features are believed to arise 
from the commensurate $(\pi,\pi)$ SDW order over a wide range of electron doping, 
as has been detected by various techniques \cite{Luke1990,Mang2004,Li2005}. A 
possible critical doping for the SDW quantum critical point (QCP) has been inferred from transport 
measurements in the normal state, which show rapidly changing transport 
properties at $x_c=0.165$ \cite{Dagan2004}. The assumption of a Fermi surface 
reconstruction caused by SDW order has led to a qualitative consistent description of
Hall effect measurements on PCCO over a wide range of doping \cite{Lin2005}.

It is important to note that there remain ambiguities about the critical value of
doping where long-range magnetic order sets in.
Elastic neutron scattering measurements on NCCO show that long-range 
magnetic order is preempted by short ranged antiferromagnetism for dopings
between $x=0.134$ and $x=0.154$\cite{Motoyama2006}. 
It has still to be clarified whether short-ranged antiferromagnetism
below optimal doping applies also to other electron-doped materials and
is confirmed also by other techniques. We will argue that 
the main features of Fermi surface reconstruction observed in electron doped
cuprates are induced by a true SDW gap. Especially in PCCO, there
is no experimental evidence that magnetic order is short-ranged below optimal doping.

Our main result is that the related Nernst effect measurements on PCCO and NCCO can 
be explained by the emergence of hole-like carriers near optimal doping. 
These aspects will be quantified within a simple semiclassical Boltzmann approach.

\section{\label{method} Model}

We consider electrons moving on a square lattice with dispersion 
\begin{eqnarray}
\varepsilon_{\bf k} = &-& 2 t_1 (\cos k_x +\cos k_y ) + 4 t_2 \cos k_x \cos k_y \nonumber\\
               &-&2 t_3 (\cos 2 k_x +\cos 2 k_y) 
\label{dispersion}
\end{eqnarray} 
and parameters $t_1=0.38$\,eV, $t_2=0.32t_1$ and $t_3=0.5t_2$ \cite{Andersen1995},
chosen to reproduce the Fermi surface measured in photoemission experiments 
\cite{Armitage2002,Matsui2007}. We will focus on a carrier density corresponding to 
the electron-doped case, with a two-dimensional density $n=1+x>1$ per unit cell. Below 
critical doping $x_c=0.165$, we assume commensurate  SDW order at wavevector ${\bf Q}=(\pi,\pi)$ 
with scattering amplitude $\Delta$ \cite{Dagan2004,Li2005}. Microscopically, this order
can be understood as a consequence of electron-electron interactions \cite{Gruener1994},
which are minimally described by the Hubbard interaction 
\begin{equation}
H_{el-el}=\frac{U}{N}\sum_{{\bf k},{\bf k^\prime},{\bf q},\sigma} c_{{\bf k},\sigma}^\dagger c_{{\bf k+q},\sigma}^\pdag c_{{\bf k^\prime},-\sigma}^\dagger c_{{\bf k^\prime}-{\bf q},-\sigma}^\pdag \ .
\label{hubbardu}
\end{equation}
The spin density wave instability is described by the complex order parameter
\begin{equation}
\Delta e^{i\phi}=\frac{U}{N}\frac{1}{V} \sum_{{\bf k}} \langle c_{{\bf k}\uparrow}^\dagger c_{{\bf k} + {\bf Q},\uparrow}^\pdag \rangle=\frac{U}{N}\frac{1}{V} \sum_{{\bf k}} \langle c_{{\bf k}\downarrow}^\dagger c_{{\bf k}+{\bf Q},\downarrow}^\pdag \rangle \ ,
\end{equation}
which can be determined self-consistently by employing
the Hartree-Fock decoupling $c_{{\bf k},\sigma}^\dagger c_{{\bf k}+{\bf q},\sigma}^\pdag \rightarrow \langle c_{{\bf k}\sigma}^\dagger c_{{\bf k}+{\bf Q},\sigma}^\pdag \rangle \delta_{{\bf Q},{\bf q}}$
in Eq.~(\ref{hubbardu}) \cite{footnote1}. 

\begin{figure}
\includegraphics[width=2.6cm]{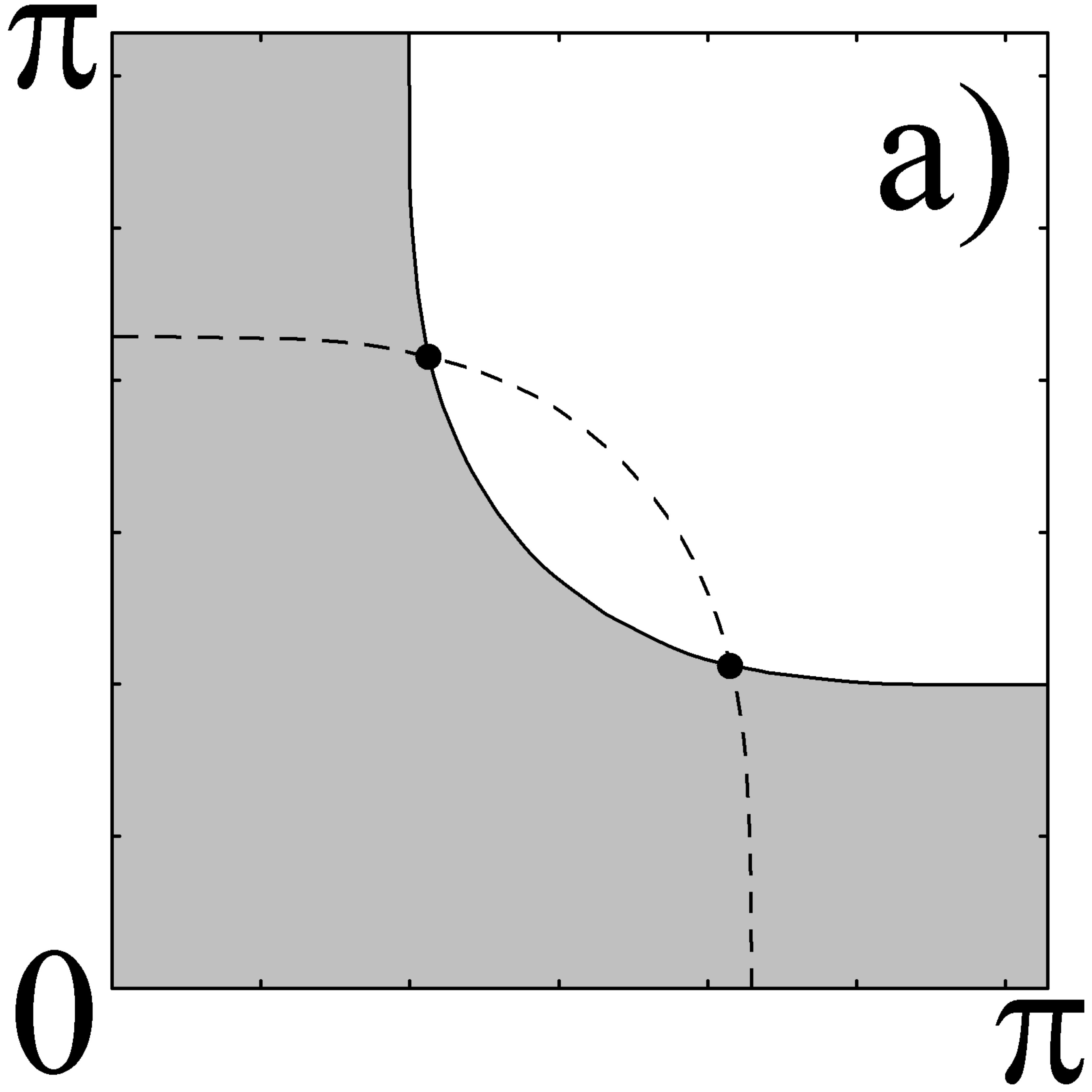} \includegraphics[width=2.6cm]{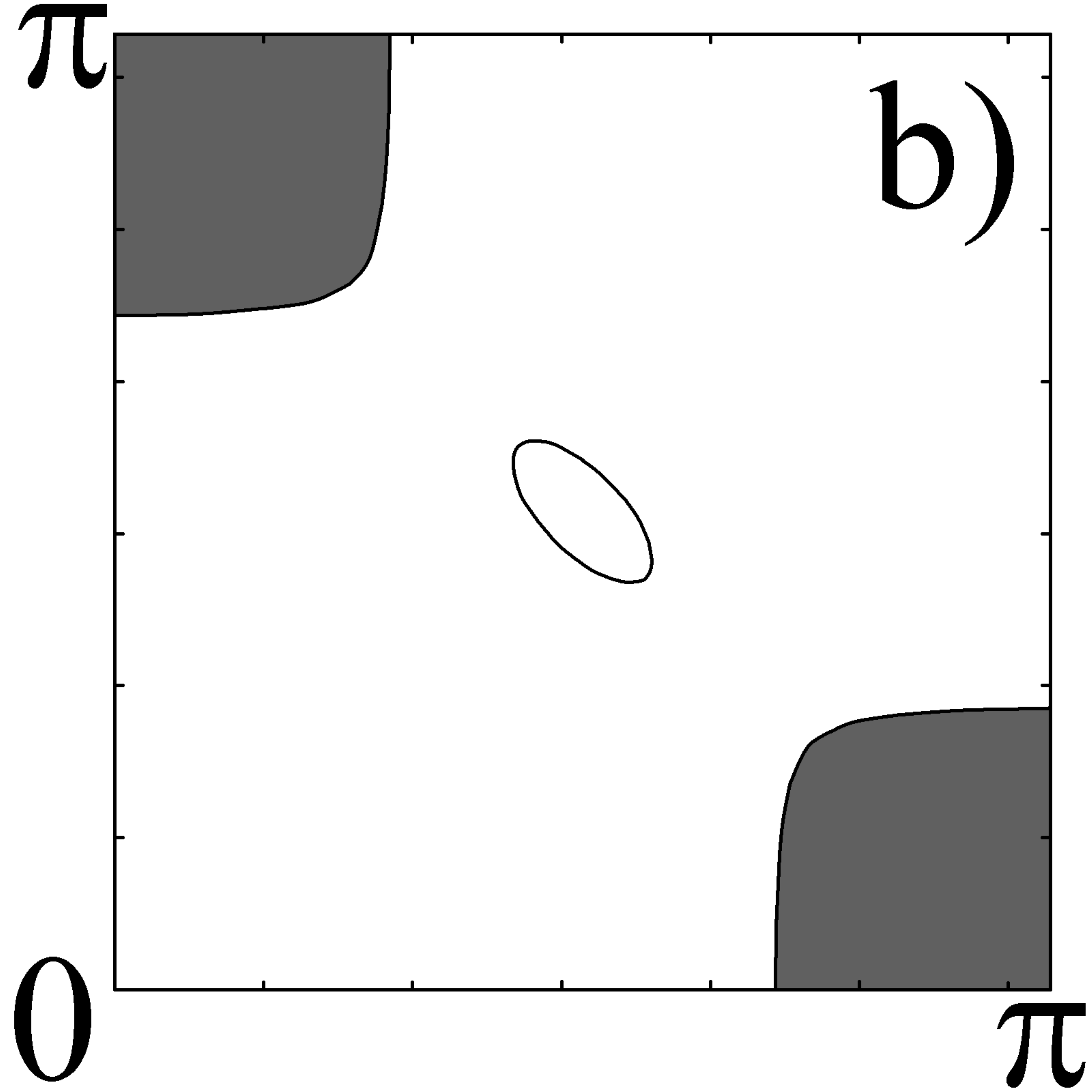} 
\includegraphics[width=2.6cm]{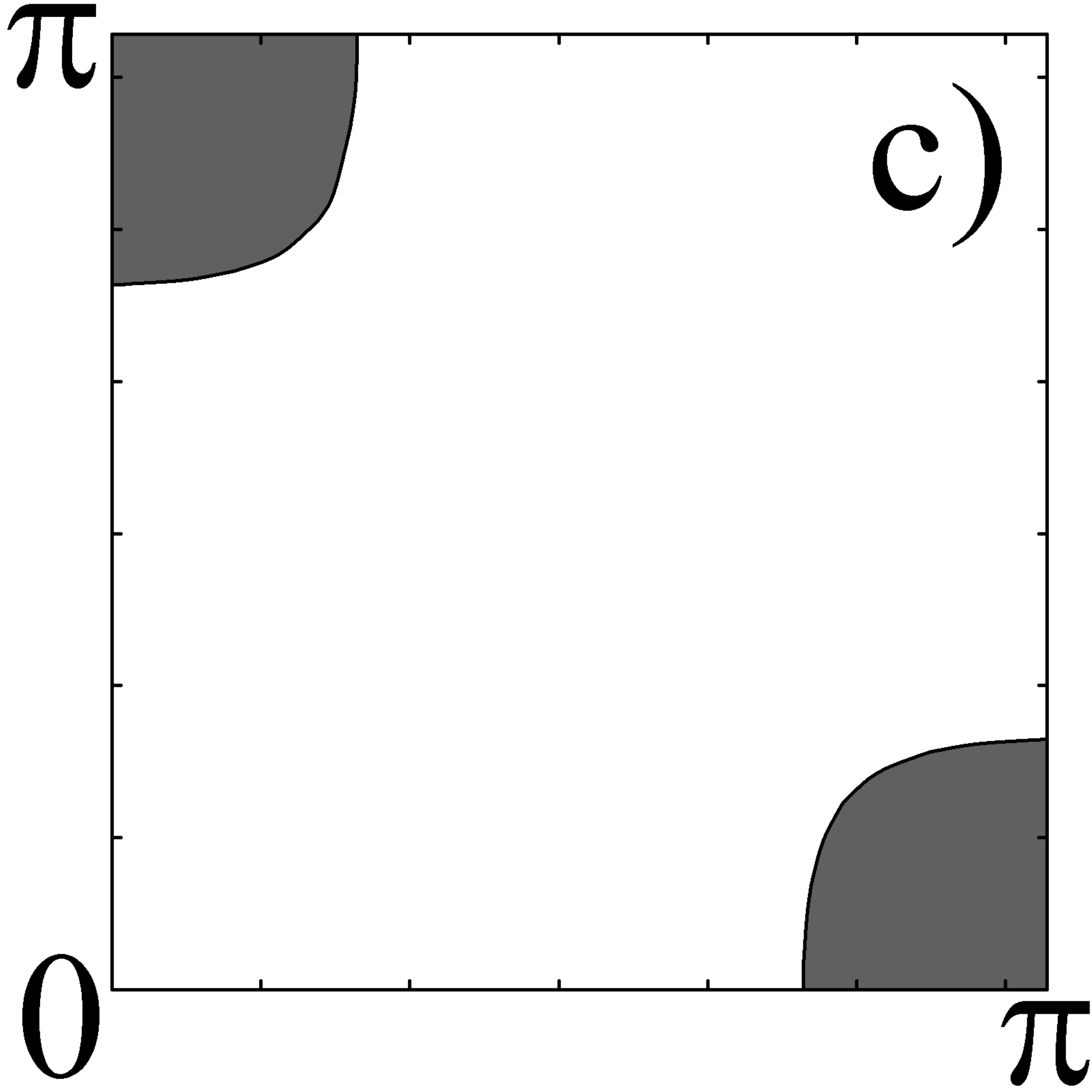} \\ 
\caption{\label{reconstruction} Evolution of the Fermi surface upon decreasing electron doping $x$. 
To distinguish holes from electrons, electrons from the upper band $E_{{\bf k}}^+$ are dark shaded, 
light shading contains all electrons from both bands. At $x=x_c=0.165$, a gap opens at the points 
where the dispersion crosses its translation by the wavevector $(\pi,\pi)$, see a). A hole pocket 
centered at $(\pi/2, \pi/2 )$ is present for $x_1 < x < x_c$ (with $x_1 = 0.145$), as shown 
in b) for $x=0.15$. For $x<x_1$, only electron-like pockets remain, as shown in c) for $x=0.12$.}
\end{figure}

In the doubled unit cell and at mean field level, this changes the dispersion to
\begin{equation}
E_{\bf k}^\pm=\frac{1}{2}\biggl( \varepsilon_{\bf k} + 
\varepsilon_{\bf k +Q} \pm \sqrt{(\varepsilon_{\bf k} - \varepsilon_{\bf k+Q})^2+4\Delta^2}\biggr) \ ,
\label{twoband}
\end{equation}
where now the reduced antiferromagnetic Brillouin zone has to be considered.
The quasiparticles resulting from the reconstructed bands have the velocities
\begin{equation}
v_{\bf{k}}^\pm=\frac{1}{\hbar} \nabla_{\bf{k}} E_{{\bf k}}^\pm/\hbar \ ,
\label{velocities}
\end{equation}
where we will omit an explicit band label from the velocities in the following 
in order to compactify our notation.
Consistent with the Hartree-Fock treatment of the effective Hamiltonian, we chose a 
mean field dependence $\Delta(x)[eV]=0.7\sqrt{1-x/0.165}$\,. The gap opens rapidly
on depleting the carrier concentration below $x_c=0.165$ and the Fermi surface reconstructs 
in qualitative agreement with ARPES data \cite{Armitage2002,Matsui2007}, see Fig.~\ref{reconstruction}.

A gap of $\Delta=0.7$\,eV yields also consistent results for the Hall coefficient \cite{Lin2005}. Our 
results are not sensitive to precise parameter choices, and slight variations of parameters lead
only to minor modifications of our results. We will show that the opening of a hole pocket will strongly 
influence the Nernst effect. Moreover, our modeling agrees with Hall measurements by Onose et al. 
\cite{Onose2001}, which indicate that the hole pockets are present for $x_1 < x < x_c$ Ce doping with  $x_1=0.1$.

\section{\label{semiclassical} Semiclassical approach}

Several parameter scales have to be set to justify our Boltzmann approach.
Backscattering of the SDW amplitude sets a momentum scale $p_\Delta=\Delta /v_F$ ($v_F$ is the Fermi velocity), 
while the inverse mean free path $l^{-1}$ defines another momentum scale.
To neglect interference effects between scattering events, the 
momentum scale $p_0$ set by the size of the Brillouin zone has to fulfill 
$p_0 \gg p_\Delta,l^{-1}~$ \cite{Lin2005}. At low $T$, we assume that impurity scattering 
dominates the relaxation time $\tau$. In general, cuprate materials show a normal state
quasiparticle scattering rate which is linear in temperature, with a small
part of the antinodal region where there is no temperature dependence observed \cite{Valla2000}.
We will neglect effects of anisotropy and temperature on the scattering rate by assuming pure s-wave 
impurity scattering, as is appropriate
for randomly distributed impurities with weak and short-ranged scattering potential.
Lateron, we will return to possible modifictations due to scattering anisotropy 
and thermal fluctuations.

Disorder is expected to modify the SDW backscattering if the mean free path $l$
drops below the characteristic scattering length on the SDW order parameter.
This situation is expected to occur if $p_\Delta l\sim 1$, and we will consider only $\Delta>v_F/l$. 
Finally, weak magnetic fields make it possible to expand transport coefficients in magnetic field strength, so that 
off-diagonal transport coefficients become linear in $B$, while the diagonal coefficients are independent of $B$.
The applicability of this expansion is related to the momentum scale set by $a=\pi l B/\phi_0$ with the flux 
quantum $\phi_0=hc/2e$, which defines the {\it weak-field regime\/} $a<p_\Delta$ \cite{Lin2005}, where the Zener-Jones
expansion is applicable. Magnetic fields also have to be weak enough to neglect magnetic breakdown. Neglecting modifications
of magnetic field on the band structure, magnetic breakdown is analogous to Zener breakdown and has a transmission
amplitude \cite{Blount1962}
\begin{equation}
\alpha=\exp(-\frac{\pi}{2}\frac{\Delta^2}{e\hbar B|v_x v_y|})
\label{breakdown}
\end{equation}
with the Fermi velocities $v_x,v_y\approx v_F$ of the linearized dispersion at its crossing point
obtained by setting $\Delta=0$. Therefore, magnetic breakdown can be neglected as long 
as $p_\Delta>p_B$, where the inverse magnetic length $p_B=2\pi(\pi B/\Phi_0 )^{-1/2}$ appears.

In mean-field approximation, the transport processes are determined
by the current operator
\begin{equation}
{\bf j} = -e \sum_{\sigma} \int_{RBZ} \frac{d^2 {\bf k}}{(2\pi)^2} 
\psi_{{\bf k},\sigma}^\dagger \left(
\begin{array}{cc}
\nabla_{\bf k} E_{\bf k}^+/\hbar  & v_{\bf k}^{inter} \\
v_{\bf k}^{inter} & \nabla_{\bf k} E_{\bf k}^-/\hbar \\
\end{array}
\right) \psi_{{\bf k},\sigma} \ , 
\label{current}
\end{equation}
where the spinor $\psi_{{\bf k},\sigma}$ contains the two quasiparticle modes.
The current therefore includes also scattering events between bands mediated by the 
off-diagonal elements
\begin{equation}
v_{\bf k}^{inter}=-\frac{1}{\hbar}\frac{[\nabla_{\bf k} \epsilon_{\bf k} - \nabla_{\bf k} \epsilon_{\bf k+Q}]\Delta}{\sqrt{(\epsilon_{\bf k}-\epsilon_{\bf k+Q})^2+4\Delta^2}}
\end{equation}
However, if the energy gap to the second band is larger than $k_B T$ and $\hbar/\tau$, interband contributions 
to transport can be neglected. We will neglegt a small doping range very close to the QCP, where $p_\Delta$ might be small
enough to allow for magnetic breakdown or modifications due to disorder. For magnetic fields of 
order a few Tesla and scattering times of $\mathcal{O} (10^{-14}s)$, this doping range is expected to be difficult
to detect in experiment. According to formula (\ref{breakdown}), magnetic breakdown is of importance in the doping range 
$\Delta x \approx e\hbar v_F^2 B / (0.7eV)^2 x_c \approx 7.1 \times 10^{-5} B x_c$, where
we used the universal Fermi velocity $v_F=2.3 \times 10^7cm/s$ \cite{Zhou2003}. Interband transitions
mediated by impurity scattering are estimated to occur in the doping range 
\begin{equation}
\Delta x \approx \biggl(\frac{\hbar}{\tau}\biggr)^2 \frac{x_c}{(0.7eV)^2} \approx 8.6 \times 10^{-3} \biggl( \frac{10^{-14}s}{\tau} \biggr)^2 x_c \ ,
\end{equation}
which is negligible for relaxation times of  $\mathcal{O}( 10^{-14} s)$. From experimental data at optimal doping, 
the relaxation time can be estimated to be somewhat larger than $10^{-14}s$, see below.
Assuming an ordering temperature $T_{SDW}=T_0\sqrt{1-x/x_c}$ with $T_0\approx 250K$ \cite{Yu2007}, 
thermal excitations destroy the SDW gap in a range of width $\Delta x \approx x_c (T/T_0)^2$ below doping $x_c$.
Keeping this in mind, we assume that all mentioned considerations are valid for 
the parameter regimes discussed below.

We define the thermoelectric response in the absence of an electrical current as 
\begin{equation}
{\bf E}=- \hat{\vartheta}  { \vec{\bf \nabla}} T \ ,
\end{equation}
from which the Nernst signal $e_N=\vartheta_{yx}$ and the thermoelectric power $Q=\vartheta_{xx}$ are
obtained. For square lattice geometry, the diagonal entries of all transport tensors
are isotropic. Both coefficients can be expressed as
\begin{eqnarray}
\vartheta_{yx}&=&\frac{\alpha_{xy}\sigma_{xx}-\alpha_{xx}\sigma_{xy}}{\sigma_{xx}^2+\sigma_{xy}^2} \nonumber\\
\vartheta_{xx}&=&\frac{\alpha_{xx}}{\sigma_{xx}} \ ,
\label{decomposition}
\end{eqnarray}
where the usual definitions of the electrical and thermoelectrical conductivities enter 
\cite{Ziman}. To calculate the quasiparticle Nernst signal, we restrict us to the weak-field
regime defined above. From the linearized Boltzmann equation,
we obtain the transport coefficients \cite{Ziman}
\begin{eqnarray}
\alpha_{xx} &=&  \frac{2e}{T} \sum_{{\bf k}, \alpha= \pm} \frac{\partial f_{\bf k}^0}{\partial E_{\bf k}^\alpha} 
(E_{\bf k}^\alpha -\mu ) \tau_{\bf k} (v_{\bf k}^x)^2    \nonumber\\
\alpha_{xy} &=&  \frac{2e^2B}{T\hbar c} \sum_{{\bf k}, \pm} \frac{\partial f_{\bf k}^0}{\partial E_{\bf k}^\alpha} 
(E_{\bf k}^\alpha-\mu ) \tau_{\bf k}^2 v_{\bf k}^x   
\biggl[ v_{\bf k}^y \frac{\partial v_{\bf k}^y}{\partial k_x} -  v_{\bf k}^x \frac{\partial v_{\bf k}^y}{\partial k_y}\biggr]\nonumber\\   
\sigma_{xx} &=&  -2e^2 \sum_{{\bf k}, \pm} \frac{\partial f_{\bf k}^0}{\partial E_{\bf k}^\alpha} \tau_{\bf k} (v_{\bf k}^x)^2  \nonumber\\
\sigma_{xy} &=& -2 \frac{e^3 B}{\hbar c} \sum_{{\bf k}, \pm} \frac{\partial f_{\bf k}^0}{\partial E_{\bf k}^\alpha} 
\tau_{\bf k}^2 v_{\bf k}^x  \biggl[ v_{\bf k}^y \frac{\partial v_{\bf k}^y}{\partial k_x} -  v_{\bf k}^x \frac{\partial v_{\bf k}^y}{\partial k_y}\biggr] \ ,
\label{boltzmann}
\end{eqnarray}
where $\alpha=\pm$ denotes summation over the quasiparticle bands of Eq.~(\ref{twoband}). For brevity, we have 
droped the band index from the quasiparticle velocities, which have been properly defined in Eq.~(\ref{velocities}).
It will be of interest to study Eq.~(\ref{boltzmann}) in dependence of electron doping in order to 
analyze the influences of Fermi surface changes on transport properties. At low
$T$, the thermoelectric conductivities $\alpha_{ij}$ are related to the electrical conducticities
$\sigma_{ij}$ by the Mott relation
\begin{equation}
\alpha_{ij}=-\frac{\pi^2}{3} \frac{k_B^2T}{e} \frac{\partial \sigma_{ij}}{\partial \mu} \biggl|_{E_F} \ .
\label{linear}
\end{equation}
As long as the relaxation time depends on energy, the expression 
$\frac{\partial \sigma_{ij}}{\partial \mu} \biggl|_{E_F}$ contains a contribution 
\begin{equation}
\frac{\frac{\partial \tau}{\partial \mu}|_{E_F} }{\tau}  (2-\delta_{ij})\sigma_{ij} \ .
\label{energy}
\end{equation}

\subsection{Comparison with experiment}

Due to Eq.~(\ref{linear}), the energy dependence $\partial \tau / \partial \mu$ enters thermoelectric quantities.
We rule these contributions out by using a constant $\tau$, in order to focus on the role
of Fermi surface geometry in the Nernst effect. Usually, the energy dependence of $\tau$ 
is expected to behave as $\tau \propto E^p$, with $p \in [-1/2,3/2]$ \cite{Barnard}. 
In the low temperature regime, according to Fermi's golden rule $\tau \propto 1/N(\epsilon)$, with $p=0$
for the two dimensional Fermi gas. Phonon contributions become only of importance for $T \gtrapprox \Theta_D$, for which 
p=3/2. Thus, phonons lead to a positive contribution in Eq.~(\ref{energy}). In two dimensions,
energy dependence of the relaxation time due to impurities yields corrections 
to the Nernst signal which vanish in the free electron case, making them sensitively dependent on
details of the band structure.

\begin{figure}
\includegraphics[width=7.5cm]{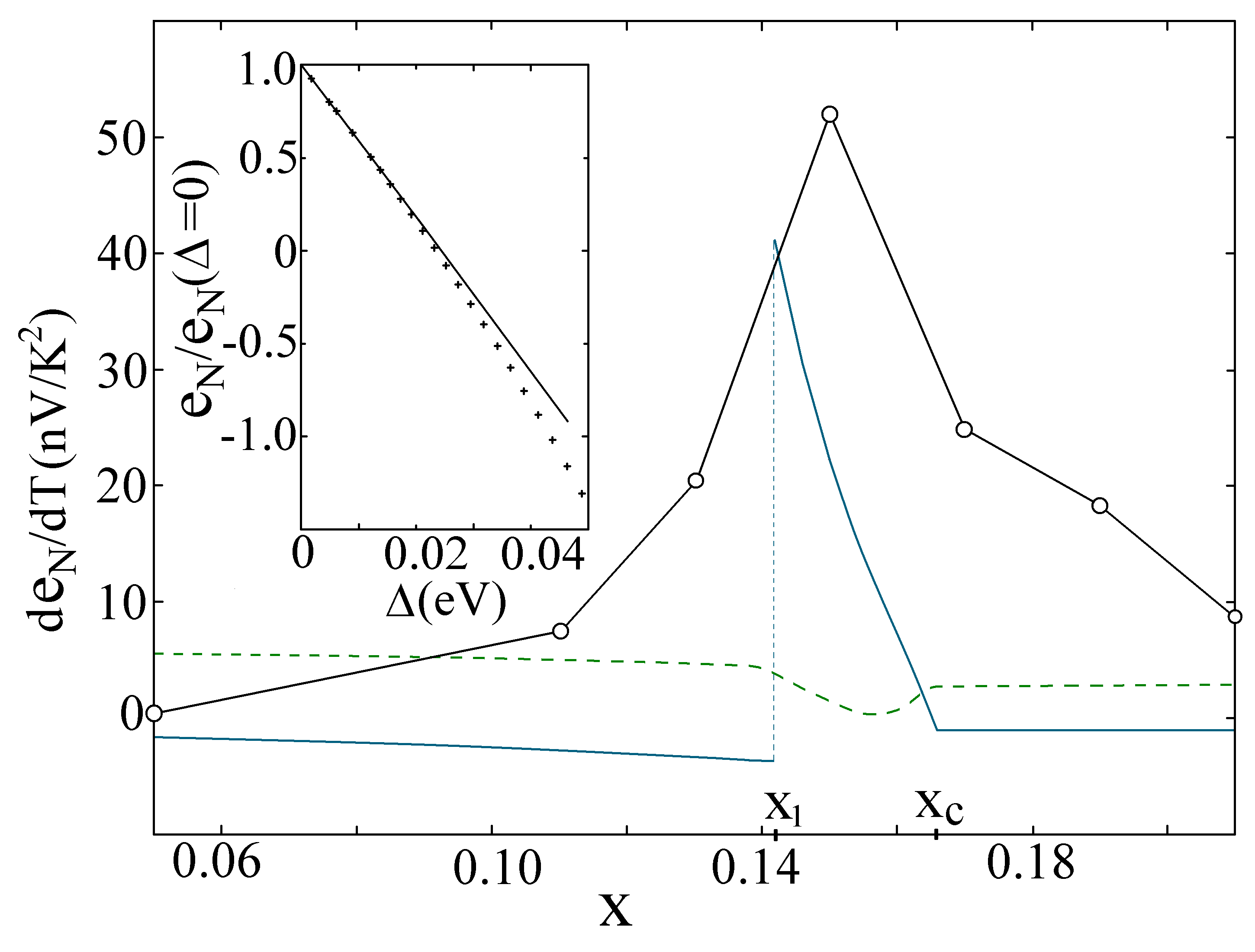}
\caption{\label{nernst} Dependence of the Nernst coefficient on electron doping in the limit $T\rightarrow 0$.
With decreasing $x$, the coefficient has an onset near $x=x_c$, where SDW order sets in;
The discontinuity at $x=x_1$ is due to the opening of hole pockets (blue curve). 
The magnitude of our estimate of contributions due to energy dependence of the 
relaxation time has negligible size in the peak region (dashed line), as
compared to the experimental values (black curve). Experimental data points  
from Ref. \onlinecite{Greene2007} correspond to
the small circles, the line is a guide to the eye.
The inset shows the quantum critical contribution to $\vartheta_{yx}$, which becomes 
large already at small gap energies $\Delta$. Numerical data points in the inset correspond to the crosses,
which asymptotically behave as a linear function of gap amplitude, as given by the black line.
}
\end{figure}

We estimated these effects numerically by setting $\tau^\prime \equiv (\partial \tau / \partial \mu)_{E_F}=\tau / E_F$, which 
yields a negligible correction to the peak signal, see Fig.~\ref{nernst}. On the 
other hand, if $\partial \tau / \partial \mu$ would contribute considerably to 
the Nernst signal, employing  Eq.~(\ref{linear}) in Eq.~(\ref{decomposition}) shows that 
$\vartheta_{yx}=\mathcal{O}(\vartheta_{xx} \tan(\Theta_H))$ with $\tan(\Theta_H)= \sigma_{xy}/\sigma_{xx}$.
However, Nernst measurements on PCCO clearly show $\vartheta_{xx}\tan(\Theta_H)\ll \vartheta_{yx}$ 
for all Ce concentrations $x>0.05$ \cite{Greene2007}, and we can thus neglect $\partial \tau / \partial \mu$.

We solved Eq.~(\ref{boltzmann}) numerically in the regime where $\vartheta_{yx}$ and $\vartheta_{xx}$ 
depend linearly on $T$, as shown in Fig.~\ref{nernst}. The experimental peak height near
optimal doping is reproduced in order of magnitude by the experimental value
$\tau=3.30 \times 10^{-14}s^{-1}$ at optimal doping, which is obtained from the residual ab-plane 
resistivity $\rho=57\mu\Omega$\,cm \cite{Dagan2004} and the plasma frequency 
$\omega_p=13000$\,cm$^{-1}~$ \cite{Homes2006}. In a range above optimal doping, the peak structure of 
the experimental signal is comparable with our theory. The experimental Nernst signal seems to be 
shifted by $\Delta x \approx 0.02$ on the doping axis, suggesting that the carrier concentration of the sample
differs from nominal doping by the same amount. A deviation of $2\%$ carrier concentration is
quantitatively also found in a comparison of the Fermi volume found from ARPES and the Fermi volume
calculated from Eq.~(\ref{dispersion}) \cite{Millis2005}. In addition, a calculation of
the Hall coefficient in dependence of electron doping using the dispersion of Eq.~(\ref{dispersion})
shows also a shift of about $2\%$ carrier concentration with respect to experimental
results in the underdoped regime, which also fail to reproduce the expected $R_H\propto 1/x$ 
behavior if $x$ is set equal to the Ce concentration \cite{Lin2005}.
The deviation could be caused by high $T$ oxygen annealing, which leads to
doping inhomogeneity/uncertainty in large crystals \cite{Kang2005}.

We therefore interpret the peak in the Nernst measurements
near optimal doping as a result of an emerging hole pocket.
A related enhancement of the Nernst signal near van Hove singularities has been 
described by Livanov \cite{Livanov1999}. The Nernst signal further away from optimal 
doping is not accurately reproduced by our model; anisotropy of the scattering rate 
\cite{Valla2000} is a possible origin of the sizable signal,
and scattering off order parameter fluctuations
should also be considered \cite{Lin2005}.

The relaxation time approximation (RTA) is a doubtful 
method to reproduce the influence of antiferromagnetic fluctuations on transport
properties. In RTA, the quasiparticle current is given by $J_{\bf k}=\tau_{\bf k} v_{\bf k}$, thereby neglecting
current vertex corrections $\Delta J_{\bf k}$ caused by the interaction-induced drag of
surrounding quasiparticles. These corrections are important to maintain a conserving approximation
in the sense of Kadanoff and Baym \cite{Kadanoff1961}. The influence of antiferromagnetic fluctuations 
on the Nernst signal is more accurately treated within the FLEX+t-matrix approximation\cite{Kontani2005}, 
which is beyond the scope of this paper. We will analyze corrections due to antiferromagnetic fluctuations
at low temperatures more detailed in section \ref{fluctuations}.

\subsection{Nernst effect near singular doping}

The behavior of the Nernst coefficient near the singular dopings in Fig.~\ref{nernst} can be obtained from 
analytical considerations. First, we consider the singularity at doping $x_1$, where 
hole like carriers emerge in the Fermi volume, see Fig. \ref{reconstruction}. The hole pocket corresponds to 
a local minimum of the dispersions Eq.~(\ref{twoband}), whose distance from the chemical potential can be expanded as 
$\Delta E = (d\mu/dx )_{x_1} (x-x_1)+\mathcal{O}(x-x_1)^2$, as we confirmed numerically. Analogous to the discussion in the 
context of the SDW gap, the gap energy $\Delta E$ has to be large enough in order to neglect magnetic breakdown and thermal excitations
across the gap. These effects tend to smear out the discontinuity in the Nernst signal over a finite range of doping,
while the order of magnitude in change in the signal is not expected to change considerably.

The asymptotic low temperature limit of the Sommerfeld expansion is valid as long as $k_B T \ll | \Delta E |$,
and thus thermal excitations change the behavior near the hole pocket in a finite range 
$\Delta x \approx (k_B T)/ |d\mu/dx|_{x_1}$ of doping. Considering the numerical value
$d\mu / dx|_{x_1} \approx 1.52 eV$, $\Delta x<0.01$  at $T<100K$.
An estimate of the doping range
where magnetic breakdown according to formula (\ref{breakdown}) can occur
is given by 
\begin{equation}
\Delta x \approx \frac{v_F}{|\frac{d\mu}{dx}|_{x_1}} \sqrt{\frac{2\hbar e}{\pi} B} \approx 5 \times 10^{-3} \sqrt{B} \ ,
\end{equation}
where the universal Fermi velocity $v_F=2.3 \times 10^7 cm/s$ and the numerical value $|\frac{d\mu}{dx}|_{x_1} \approx 1.52$ \,eV
have been used. For experimentally relevant magnetic field strengths of $B \approx 10T$, the Nernst signal
is therefore expected to become sharply enhanced already for dopings of about $1-2 \%$ below $x_1$, consistent with the experimental
result shown in Fig. \ref{nernst}.

Near the opening of the hole pocket at $x=x_1$, the hole dispersion 
is approximated by $\varepsilon_h({\bf k})= \sum_i  \delta k_i^2/m_i-\mu_h$, and 
the $T=0$ hole contributions to electrical transport become
\begin{eqnarray}
\sigma_{xx}^h(\mu_h)&=&  \frac{2}{3} \mu_h \tau_h(\mu_h) e^2 \frac{N_h}{ \bar{m}_h} \nonumber\\
\sigma_{xy}^h(\mu_h)&=&  \frac{2}{3} \mu_h \tau_h^2(\mu_h) \frac{e^3B}{c} \frac{N_h}{ \bar{m}_h} 
\label{openpocket}
\end{eqnarray}
for $\mu_h>0$ and vanish otherwise. In the following, we formally distinguish electron and hole 
scattering rates. The hole DOS $N_h$ and the reduced hole mass $\bar{m}_h=(m_1 m_2)/(m_1+m_2)$ 
are taken to be constant. For weak dilute disorder, the scattering rate follows $1/\tau(\mu_h) \propto N_h$ and 
is energy independent. According to Eqs (\ref{linear}) and (\ref{openpocket}), the Nernst signal and the thermopower 
have discontinuities at $\mu_h=0$
\begin{eqnarray}
\Delta \vartheta_{yx} &=& \biggl[ \frac{\sigma_{xx}^e \alpha_{xy}^h - \sigma_{xy}^e \alpha_{xx}^h}{(\sigma_{xx}^e)^2} \biggr]_{\mu_h=0^+}\nonumber\\
\Delta \vartheta_{xx} &=& \biggl[ \frac{\alpha_{xx}^h}{\sigma_{xx}^e} \biggr]_{\mu_h=0^+} \ .
\label{thermojump}
\end{eqnarray}
Expanding the electron dispersion as $\varepsilon({\bf k})= \sum_i \delta k_i^2/m_i-\mu$, the relative changes are 
\begin{eqnarray}
\frac{\Delta \vartheta_{yx}}{\vartheta_{yx}|_{\mu_h=0^-}}&=& -\frac{N_h \bar{m}_e \tau_h}{N_e \bar{m}_h \tau_e} \biggl[ \frac{\tau_h +\tau_e}{\tau_e^\prime \mu_e}\biggr] \nonumber\\
\frac{\Delta \vartheta_{xx}}{\vartheta_{xx}|_{\mu_h=0^-}}&=&  -\frac{\tau_h N_h \bar{m}_e}{\tau_e N_e \bar{m}_h}\ .
\label{ellipticaljump}
\end{eqnarray}
Sizable contributions from the discontinuity can therefore be expected, and $ \tau_e^\prime <0$ 
would explain why the Nernst signal shows no sign change in experiments on PCCO \cite{Greene2007}.
Moreover, a sign change near $x=0.15$ has been found in the thermoelectric power \cite{Li2005}, as 
predicted by Eq.~(\ref{ellipticaljump}). Assuming $\tau_h N_h \bar{m}_e \approx \tau_e N_e \bar{m}_h$, the 
magnitude of the discontinuity in the thermopower is about twice the magnitude of the thermoelectric power 
in the overdoped region. This relative change in thermopower is quantitatively equivalent to the change observed 
from $x=0.15$ to $x=0.16$ in the thermopower measurements from Ref. \onlinecite{Li2005}. We briefly extend
this analysis to a general two-carrier system with carrier types 1 and 2, where 
$\Delta\vartheta_{yx}=(\sigma_{xx}^{(1)} \alpha_{xy}^{(2)}-\sigma_{xy}^{(1)} \alpha_{xx}^{(2)})/(\sigma_{xx}^{(1)})^2$ right at the emergence of carrier
type 2, since Eq.~(\ref{openpocket}) leads to $\hat{\sigma}^{(2)}=0$ at the opening of a carrier pocket.
According to Eq.~(\ref{linear}) and considering positive magnetic field strenghts $B$ in the following, $\alpha_{xy}^{(2)}$ is always positive and  
$\alpha_{xx}^{(2)}$ has always the sign of $\sigma_{xy}^{(2)}$ due to Eq.~(\ref{linear}). This means that $\Delta\vartheta_{yx}$ 
is always positive if the carriers 1 and 2 have opposite charge, while $\Delta\vartheta_{yx}$ might both be 
negative or positive if carrier type 1 and 2 have the same charge. To decide on the charges of 
carriers 1 and 2, in addition the sign of the second contribution in $\Delta \vartheta_{yx}$ can be 
determined from a measurement of $\Delta \vartheta_{xx}\tan(\Theta_H)$.

\subsection{Behavior near quantum critical point}

\begin{figure}
\includegraphics[width=7.5cm]{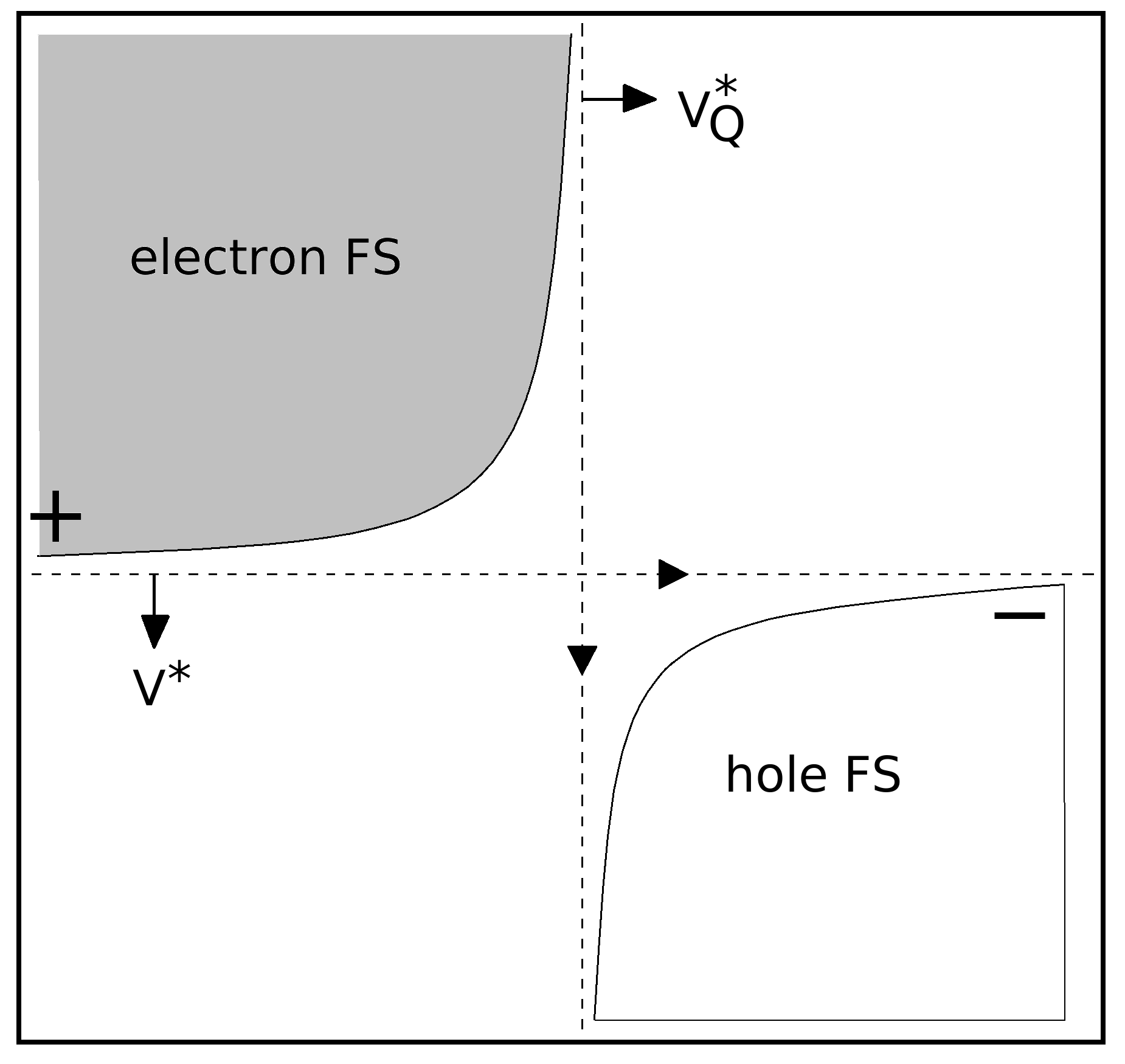}
\caption{\label{crossingpoint} To leading order in the gap amplitude $\Delta$, opening of the SDW gap modifies the Fermi surface 
only near the crossing points in momentum space where $\varepsilon_{\bf p}=\varepsilon_{\bf p+Q}=\mu$.
As shown in this sketch, a crossing point is coinciding with the crossing of the dashed lines as long
as curvature of the Fermi surfaces is neglected near the crossing point.
The vertical dashed line is the Fermi line for the normal state which is parallel to the vector $(0,\pi)$.
The horizontal dashed line is the normal state Fermi line shifted by ${\bf Q}=(\pi,\pi)$, 
thus directing parallel to $(\pi,0)$. The reconstructed Fermi surface contains electron pockets,
denoted by +, and hole pockets, denoted by -.
}
\end{figure}

We now analyze the onset of the Nernst signal at the $x=x_c$ QCP where $\Delta$ first becomes non-zero 
with decreasing $x$.
A calculation analogous to Refs \onlinecite{Bazaliy2004,Lin2005} can be employed to calculate the change 
$\delta \vartheta=\vartheta(\Delta)-\vartheta(\Delta=0)$ to linear order in the gap $\Delta$.
The changes of the dispersion to leading order in $\Delta$ occur around momenta $\bf{p}$
with $\epsilon_{\bf p+Q}=\epsilon_{\bf p}=\mu$,
which are given by the crossing points in Fig.~\ref{reconstruction}a and their symmetry related counterparts.
It is useful to parametrize ${\bf p}$ by $\varepsilon_{\bf p}$ and $\varepsilon_{\bf p+Q}$,
what is possible in the vicinity of any crossing point ${\bf p}^\star$.
This can be achieved by expanding the dispersions 
\begin{eqnarray}
\varepsilon_{\bf p}-\mu     &=& { \bf v^\star \cdot  \delta  p} + \frac{m_{ij}}{2}  \delta p_i \delta p_j \nonumber\\
                           &+& \frac{y_{ijk}}{6} \delta p_i \delta p_j \delta p_k+ \mathcal{O}(\delta p^4) \nonumber\\
\varepsilon_{\bf p +Q}-\mu  &=& { \bf v_Q^\star \cdot  \delta  p} + \frac{n_{ij}}{2}  \delta p_i \delta p_j \nonumber\\
                           &+& \frac{z_{ijk}}{6} \delta p_i \delta p_j \delta p_k+ \mathcal{O}(\delta p^4) \ ,
\label{expansion}
\end{eqnarray}
where 
\begin{eqnarray}
\delta {\bf p}&=& {\bf p} - {\bf p^\star} \nonumber\\
{\bf v^\star}&=&{\bf v(p^\star)}, ~~ {\bf v^\star_Q}={\bf v({\bf p^\star+Q)}} \nonumber\\
m_{ij}&=&(\partial^2\varepsilon_{\bf p}/\partial p_i\partial p_j)|_{{\bf p^\star}} \nonumber\\
n_{ij}&=&(\partial^2\varepsilon_{\bf p}/\partial p_i\partial p_j)|_{{\bf p^\star+Q}} \nonumber\\
y_{ijk}&=&(\partial^3\varepsilon_{\bf p}/\partial p_i\partial p_j\partial p_k)|_{{\bf p^\star}} \nonumber\\
z_{ijk}&=&(\partial^3\varepsilon_{\bf p}/\partial p_i\partial p_j\partial p_k)|_{{\bf p^\star+Q}} \ .
\end{eqnarray}
Equation (\ref{expansion}) can be inverted to yield
\begin{equation}
\delta {\bf p} = {\bf u}_1 \varepsilon_{\bf p} + {\bf u}_2 \varepsilon_{{\bf p+Q}}=({\bf u}_1 + \frac{\Delta^2}{\varepsilon_{ \bf p}^2} {\bf u}_2 ) \varepsilon_{\bf p}
\label{coordinates}
\end{equation}
with
\begin{eqnarray}
{\bf u_1} &=& {\bf \frac{v_Q^\star \times [v^\star \times v_Q^\star]}{(v^\star\times v_Q^\star)^2}} \nonumber\\
{\bf u_2} &=& {\bf \frac{v^\star \times [v_Q^\star \times v^\star]}{(v^\star\times v_Q^\star)^2}} \ .
\end{eqnarray}
Differentiating Eq.~(\ref{expansion}), substituting Eq.~(\ref{coordinates}) into it and using the result in 
Eq.~(\ref{boltzmann}) for the electrical conductivities, we obtain the linearized $T=0$ change in the electrical conductivity 
tensor $\delta {\hat \sigma} ={ \hat \sigma}(\Delta)-{\hat \sigma}(\Delta=0)$ in multiples of the conductance
quantum $\sigma_Q=e^2/\hbar$ as
\begin{eqnarray}
\delta \sigma_{xy} &=& \sigma_Q \tau^2 \frac{B \Delta}{\Phi_0 } {\bf \hat{z} \cdot [\eta_1^p + \eta_2^{sp}
+}3{\bf \eta_2^p +3\eta_1^{sp}}  ] \times ({\bf v_Q^\star -v^\star} ) \nonumber\\
\delta \sigma_{xx} &=& -\sigma_Q \frac{\tau}{\pi}\frac{({\bf v^\star - v_Q^{\star }})^2}{|{\bf v_Q^\star} \times {\bf v_Q^{\star}}|} \Delta  \ .\nonumber\\
\label{shifts}
\end{eqnarray}
Here, the vectors 
\begin{eqnarray}
{ \bf \eta^p_1}     &=& (m_{11}u_{1x} + m_{12}u_{1y}) {\hat{\bf x}} +  (m_{21}u_{1x} + m_{22}u_{1y}) {\hat{\bf y}}\nonumber\\
{ \bf \eta^p_2}     &=& (m_{11}u_{2x} + m_{12}u_{2y}) {\hat{\bf x}} +  (m_{21}u_{1x} + m_{22}u_{2y}) {\hat{\bf y}}\nonumber\\
{ \bf \eta^{p+Q}_1} &=& (n_{11}u_{1x} + n_{12}u_{1y}) {\hat{\bf x}} +  (n_{21}u_{1x} + n_{22}u_{1y}) {\hat{\bf y}}\nonumber\\
{ \bf \eta^{p+Q}_2} &=& (n_{11}u_{2x} + n_{12}u_{2y}) {\hat{\bf x}} +  (n_{21}u_{1x} + n_{22}u_{2y}) {\hat{\bf y}}\nonumber\\
\end{eqnarray}
have been defined. Parenthetically, we note that the linearized change $\delta \sigma_{xx}$ in the electrical conductivity has been 
treated in great detail previously for the three dimensional SDW transition in Cr, with essentially the same result \cite{Bazaliy2004}. 
Via Eq.~(\ref{linear}), changes in the thermoelectric conductivities are obtained from
$\frac{d \delta \sigma_{ij}}{d \mu}$. These derivatives of Eq.~(\ref{shifts}) are
obtained from the relations
\begin{eqnarray}
\frac{d  v_i^\star}{d \mu} &=& \sum_j m_{ij}( u_1^j+ u_2^j) \nonumber\\
\frac{d m_{ij}}{d \mu} &=&  \sum_k v_{ijk} (u_1^k + u_2^k) \ ,
\label{derivformulas}
\end{eqnarray}
Linearizing Eq.~(\ref{decomposition}) in $\Delta$ in this way yields $\delta \vartheta_{xx}$ and 
$\delta \vartheta_{yx}$ to linear order in $\Delta$. From a numerical calculation of $\vartheta_{xx}$
and $\vartheta_{yx}$, we obtain the values $\delta \vartheta_{xx} / \vartheta_{xx}= 47.4 \Delta $ and 
$\delta \vartheta_{yx} / \vartheta_{yx}= -39.8 \Delta$, see also Fig.~\ref{nernst}. Very close to 
$x_c=0.165$ it might be difficult to measure the quantum critical contributions $\delta \vartheta_{yx}$ 
and $\delta \vartheta_{xx}$ experimentally due to other contributions to the signal which we could 
not specify.

\subsection{Finite temperatures}

\begin{figure}
\includegraphics[width=7.5cm]{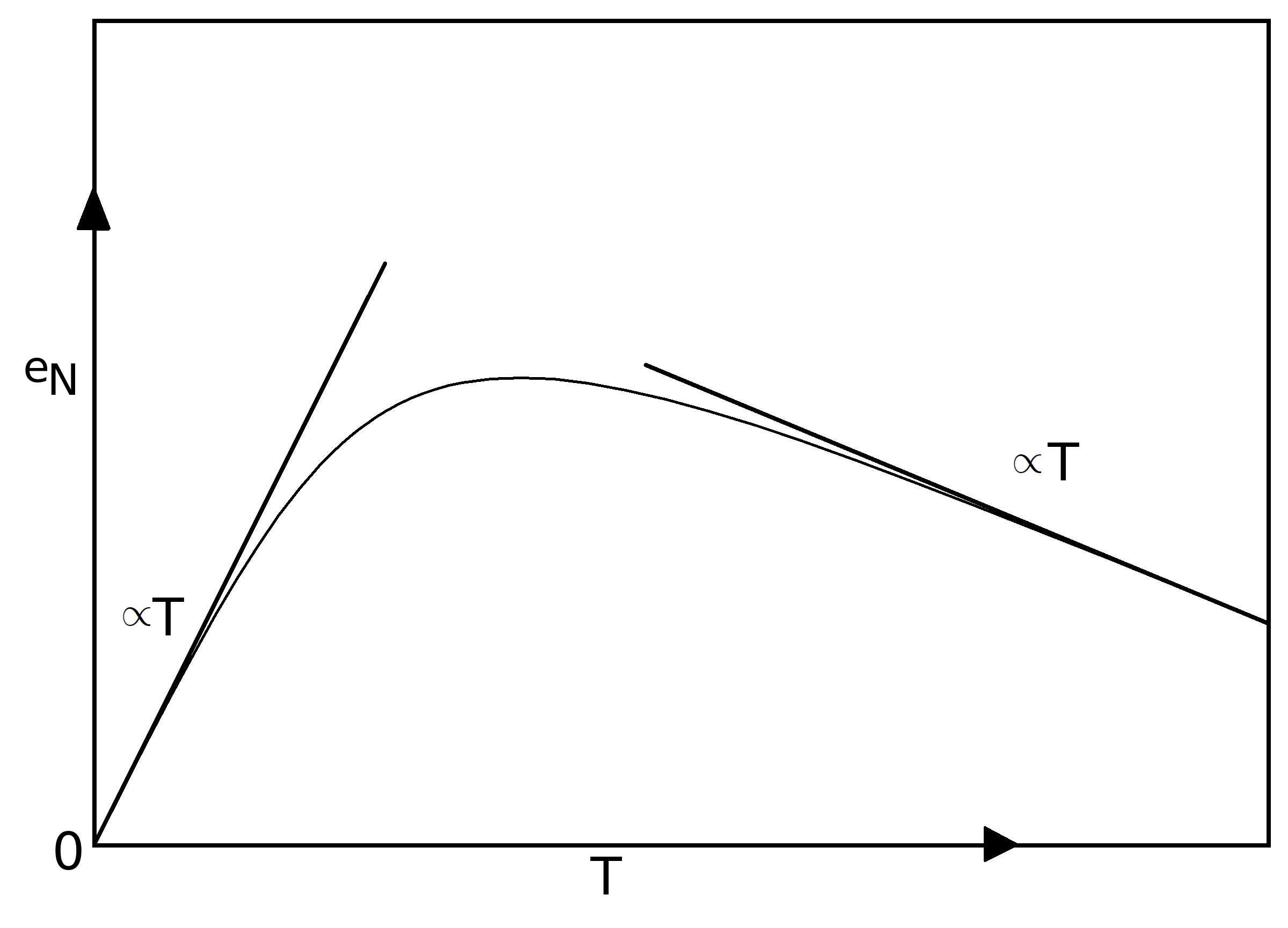}
\caption{\label{nernsttemperature} Sketch of the normal state Nernst signal dependence on temperature.
The linear temperature dependence at lowest $T$ turns over in a maximum at temperatures below
the spin density wave ordering temperature. At optimal doping, the position of this maximum is roughly $50K$.
Above the peak temperature, the signal vanishes proportional to temperature.
}
\end{figure}

At finite temperatures the Fermi surface as well as the quasiparticle 
scattering rate is expected to change. These effects will influence the 
temperature dependence of the Nernst signal which is sketched in Fig. \ref{nernsttemperature}.
First of all, in cuprate materials it has to be considered that the quasiparticle scattering rate 
is linear in temperature in most parts of the Brillouin zone \cite{Valla2000}. Important 
changes in the Fermi surface have to be considered at temperatures above $T_{SDW}$. 
Below this temperature, fluctuations of the SDW order parameter
remain gapped and can be neglected. Once the SDW gap closes, the Fermi surface reconstructs
and fluctuations of the SDW order parameter provide an important scattering 
mechanism. In the far underdoped region of electron doped cuprates, $T_{SDW}$ is of 
order the Debye temperature and scattering off phonons has to be considered as well.

In a range of temperatures above $T_{SDW}$, a sizable Nernst effect is still
observed in experiment \cite{Greene2007}. Our present mean-field theory for
Fermi surface reconstruction cannot account for the size of the signals.
However, there is no good theory for transport in this fluctuation regime.
It would be interesting to examine the behavior in a recent theory of
thermal fluctuations in the orientation of the SDW order \cite{vg}.
The experimental observations at finite temperatures 
are not in contrast with our assumptions, which are only valid at
temperatures below $T_{SDW}$, where fluctuations of the SDW order parameter are
negligible. In fact, it was shown that antiferromagnetic fluctuations enhance the Nernst signal
at finite temperatures above the magnetic ordering temperature $T_{SDW}$ and reproduce the peak structure 
in the normal state Nernst signal observed in experiment \cite{Kontani2005}. The relaxation time approximation
(RTA) certainly fails to reproduce this effect, because it would 
predict a small Nernst signal above $T_{SDW}$. Above this temperature, the normal state quasiparticles
are gapless and form a single carrier system. Within the relaxation time approximation, the Nernst signal is 
thus diminished by the Sondheimer cancellation \cite{Behnia2009}, in contrast to experimental results.

\section{\label{fluctuations} Antiferromagnetic fluctuations}

Within our assumptions, the SDW gap vanishes at a
quantum critical point upon doping with electron carriers.
There has been substantial disagreement over the position
of this quantum critical point. Elastic neutron scattering
measurements on NCCO suggest short ranged antiferromagnetic order 
between $x=0.145$ and $x=0.154$ \cite{Motoyama2006}, and it has been proposed that short ranged order
might even occur at $x=0.134$ \cite{footnote2}. Transport measurements on PCCO 
show rapidly changing transport properties  at dopings below 
$x=0.165$, suggesting that Fermi surface properties
change drastically already slightly above optimal doping.
Part of the confusion might originate from the uncertainty 
about the oxygen content of the samples, which makes it
difficult to compare the effective carrier concentration of different samples.

Experimental results show a strong doping dependence of Hall and Nernst effect way
above optimal doping \cite{Dagan2004,Greene2007}, where clearly no spin density wave gap exists.
This suggests that the band structure parameters change upon electron doping.
One way to understand this behavior is to analyze self-energy corrections
originating from antiferromagnetic spin-fluctuations. This analysis 
also helps to clarify whether short-ranged magnetic order can account
for the observed Fermi surface reconstruction and enhancement of the normal
state Nernst signal for electron dopings below $x\approx0.16$.

The effect of spin fluctuations on the Fermi surface can be obtained from 
the real part of the electronic self energy. We neglect the
imaginary part of the self energy by assuming again that impurity scattering
dominates transport at lowest temperatures. As discussed in section \ref{semiclassical},
scattering on spin flucuations would be beyond the scope of our approach 
due to the failure of the relaxation time approximation to treat this effect.
The leading approximation to the electronic self energy due to spin fluctuations is
\begin{equation}
\Sigma({\bf k},i\omega)=-g^2 T\int d^2{\bf q } \sum_{i\Omega_n} G({\bf k+q},i\omega+i\Omega_n) D({\bf q},i\Omega_n) \ ,
\end{equation} 
where $G$ and $D$ are electron and spin fluctuation Matsubara Green's functions, respectively.
We will use: $G({\bf p},i\omega)=(i\omega-\zeta_{\bf p})^{-1}$ and 
$D({\bf q},i\Omega_n)=-(\Gamma_{\bf q}+|\Omega_n|)^{-1}$, with $\zeta_{\bf p}=\varepsilon_{\bf p}-\mu$ and $\Gamma_{\bf q}=\Gamma (r+\xi^2({\bf q}-{\bf Q})^2)$,
where $\Gamma$ is the energy scale characteristic of spin fluctuations and $\xi$ the correlation length of magnetic order.
The distance to the SDW quantum critical point is controlled by the parameter $r$, and the ordering wavevector
is again ${\bf Q}=(\pi,\pi)$.

\begin{figure}
\includegraphics[width=7.5cm]{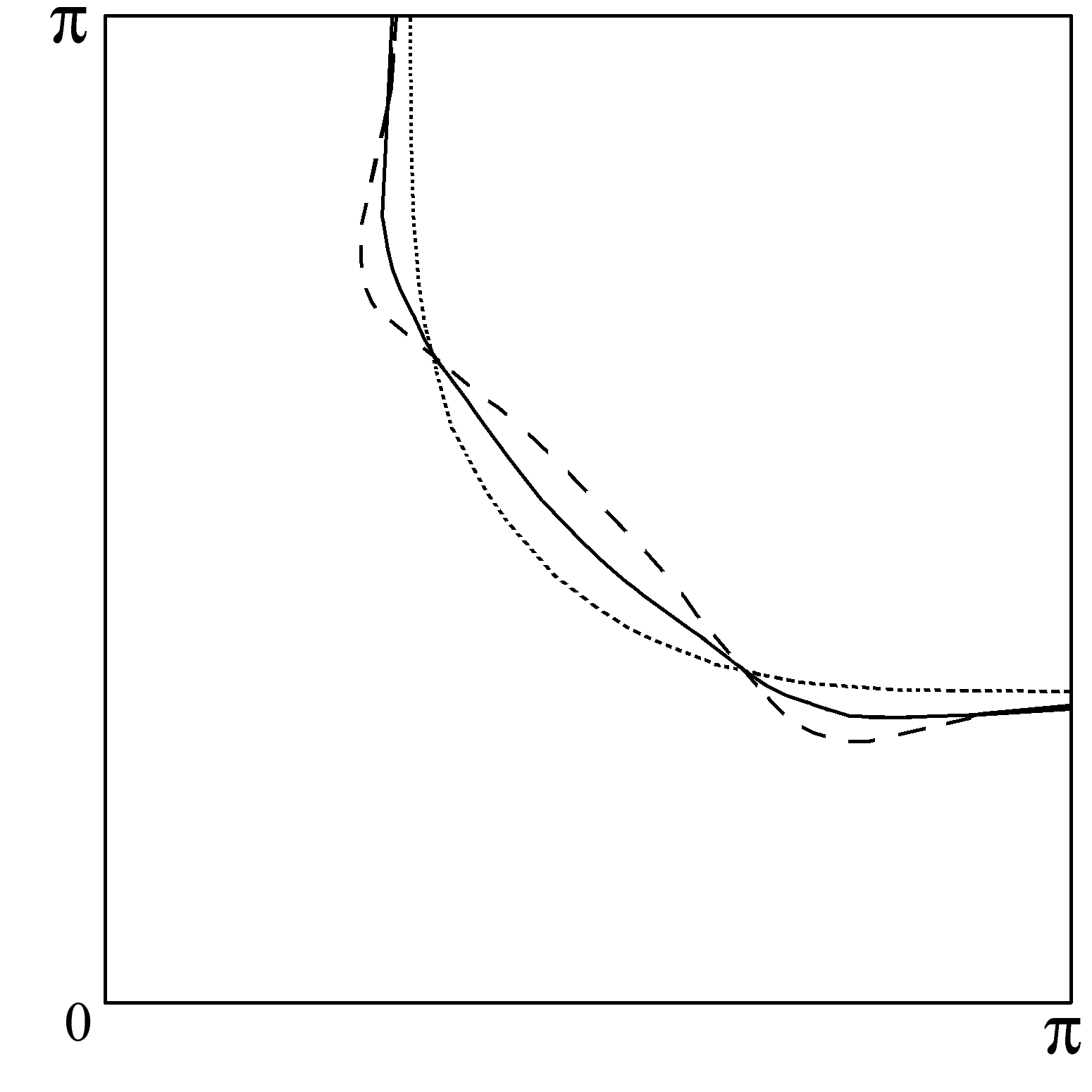}
\caption{\label{realself} Fermi surface changes due to finite spin correlation lengths
corresponding to electron dopings somewhat above the SDW quantum critical point. The Fermi surface without
any influences of spin fluctuations is shown as the dotted line.
The dashed line shows a renormalized Fermi surface for a large coupling
to spin fluctuations ($\lambda=0.4$) and a small distance $r=0.001$ to the quantum critical point.
Renormalization effects become weaker for a smaller coupling $\lambda=0.2$ and
$r=0.1$ (continuous line). At the crossing points with $\varepsilon({\bf p}+ {\bf Q})= \varepsilon({\bf p})=\mu$, 
the Fermi surface remains unchanged by spin fluctuations. The cutoff energy is $E_0=0.88$ \,eV . 
}
\end{figure}

As has been discussed in Ref. \onlinecite{Lin2005}, the self energy at $T=0$ and $i\omega=0$
can be integrated as
\begin{equation}
\Sigma({\bf k},i\omega=0)=0.5\lambda \zeta_{{\bf k}+{\bf Q}} \ln(r^2+(\zeta_{{\bf k}+{\bf Q}}/E_0)^4) \ ,
\end{equation}
where $r$ and $\zeta_{\bf{k}+{\bf Q}}/E_0$ are assumed to be small.
The energy scale $E_0$ is a cutoff of order the normal state band width and $\lambda$
is a dimensionless coupling constant. The new electronic dispersion is
\begin{equation}
\varepsilon_{\bf p}+\Sigma({\bf p},i\omega=0)\ .
\label{newdispersion}
\end{equation}
Using this new dispersion in Eqs (\ref{twoband}) and (\ref{boltzmann}), the resultant Nernst signal
(not shown) is enhanced but negative for the renormalized Fermi surfaces shown in Fig. \ref{realself}.
We obtained a negative Nernst signal for a wide range of parameters $r \in [0.001,0.1]$ and $\lambda \in [0.1,0.4]$,
including a regime of very weak spin fluctuation effects where we believe Eq.~(\ref{newdispersion}) is an accurate approximation.

We conclude that the renormalization of the Fermi surface due to spin fluctuations (states without long-range
SDW order) fails to reproduce the
Nernst signal observed in experiment. At optimal doping and below, this means that the observed Fermi
surface reconstruction is more likely to originate from long-range SDW order.
This interpretation is also supported by a comparison of the renormalized 
Fermi surfaces in Fig. \ref{realself} with ARPES measurements on NCCO \cite{Armitage2002}.
At electron dopings of $x=0.15$ and below, photoemission intensity is significantly suppressed
near $(0.65 \pi,0.3\pi)$ (and its symmetry related points) at the intersection of
the Fermi surface with the antiferromagnetic Brillouin zone boundary.
This change in photoemission intensity cannot be explained from Eq.~(\ref{newdispersion}),
since leading order self energy corrections are cancelled at any crossing point where 
$\varepsilon({\bf p}+ {\bf Q})= \varepsilon({\bf p})=\mu$. Thus again, the opening of 
a SDW gap slightly above optimal doping $x=0.15$ seems more plausible to explain the Fermi surface reconstruction 
seen in experiment. 

\section{Conclusions}

Our results show that SDW order in the electron-doped cuprates has fundamental implications 
for the Nernst signal and the thermopower. As the SDW gap becomes stronger,
the hole-like carriers will eventually vanish and the Nernst signal will have a large 
discontinuous change at the lowest $T$. This behavior is also obtained for the thermopower, 
where the discontinuity in addition should cause an observable sign change in the signal.
At finite $T$, the discontinuities will be smeared out by thermally
excited carriers and magnetic breakdown. To obtain our results, the presence of oppositely 
charged carriers represents a necessary, but not a sufficient condition in order
to obtain an enhanced Nernst signal. The fundamental origin of the maximal Nernst 
signal within our calculation is a singularity in the quasiparticle density of states, while the Nernst
signal gets weaker if the Fermi surface moves away from this singularity,
although two types of carriers are still present in the Fermi surface.
We note that the existence of oppositely charged and current carrying quasiparticles
is a widespread argument to explain an enhanced normal state Nernst signal,
but our results require a more subtle physical origin than the requirement
of two types of carriers would represent.

In this sense, our results are also in contrast with the analysis of the 
{\it ambipolar\/}  Nernst effect in Ref. \onlinecite{Oganesyan2004}, which predicts 
a maximal Nernst signal when hole and electron-like carrier densities exactly
compensate each other. This explanation had been used previously to account 
for the large normal state Nernst signal in PCCO \cite{Greene2007}. Within our
analysis, the ambipolar signal is instead largest when the hole pockets just 
touch the Fermi surface, and decreases rapidly until the carriers compensate most.

Our findings are also likely of relevance to the hole-doped cuprates.
Recent explanations of a large normal state Nernst signal in these materials 
were based on the proposal of $d$-density wave order \cite{Oganesyan2004,Tewari2009}.
A large normal state Nernst signal has recently been reported \cite{louis2009} in the 
stripe-ordered phase of La$_{1.6-x}$Nd$_{0.4}$Sr$_{x}$Cu$_4$, which vanished in the non-ordered state. These findings suggest 
that stripe order enhances the normal state Nernst effect,
and it would be interesting to extend our results to spin/charge density wave orders.

The onset of ``stripe'' order, and the evolution from ``large'' to ``small'' Fermi
surfaces with decreasing doping \cite{Choniere2009} could lead to a large Nernst signal 
by the opening/closing of hole or electron pockets. The connection of 
such normal state features to those associated with the superconductor-insulator
QPT computed earlier \cite{markus} remains an important open problem, and some ideas
have appeared in Ref.~\onlinecite{vg}.

In summary, we have presented a theory for the anomalously large normal state
Nernst signal in the electron-doped cuprates. We established a direct relation 
between SDW order and the peak of the normal state Nernst signal at optimal doping. Finally, 
while the energy dependence of the scattering rate is unlikely to modify our
result, a more detailed understanding of the scattering mechanism 
is necessary for a quantitative understanding of the large Nernst signal 
in the underdoped and overdoped regions.

We thank M.~M\"uller, V.~Galitski, R.~Greene and L.~Taillefer for useful discussions. L.~Taillefer
alerted us to observations \cite{Choniere2009} connecting stripe order in the hole-doped
cuprates to a change in the Fermi surface. 
A. H. acknowledges support by the DFG through the SFB 608 (K\"oln)
and the Research Units FG 538 and FG 960.
The research was supported by the NSF under grant DMR-0757145 and by the FQXi
foundation.

%\vskip -0.15in

\end{document}